# NEW BLACK STRINGS IN THREE DIMENSIONS

**Nemanja Kaloper** *

Department of Physics

McGill University

Montréal, Québec, Canada H3A 2T8

## Abstract

We present two different families of stationary black strings in three dimensions carrying electric and axion charges. Both solutions contain a singular region, which however is quite harmless for string frame geodesic observers, because the solutions are causally geodesically complete. In addition, we exhibit a variety of extremal limits of our solutions, and argue that there must exist a more general family continuously interpolating between them.

---

In this talk we will present two new additions to the existing bestiary of black configurations in three-dimensional string gravity[1]. We will describe how the Abelian duality symmetry is used to obtain them[2−4]. In general, this procedure for deriving new backgrounds is dubbed twisting, or $O(d,d)$ boosting, after the complete group of twisting transformations. At the level of the background field theory on target space, after integrating out Killing coordinates à la Kaluza-Klein, this is realized as a symmetry of the action under mixing of the Kaluza-Klein matter fields with the metric. Although the action is invariant under this group, solutions are not, because they employ specific initial conditions. After canceling the gauge degrees of freedom, we can see that the space of classical solutions is spanned by the orbits of the $O(d,d)$ group, modulo diffeomorphisms and Kalb-Ramond gauge transformations. This symmetry is further extended to $O(d,d+n)$ in the presence of $n$ Abelian gauge fields.

With the help of this extended boost symmetry, we have found two new three-dimensional (3D) families of asymptotically flat solutions. They have been obtained by, respectively, "twisting in" the gauge field on the black string of Horne and Horowitz[5], and the axion on the 2D electrically charged black hole[6] crossed with a flat line. They are characterized by three parameters, and represent different stationary, gauge charged configurations with regular horizons. Although all our non-extremal black strings possess a scalar curvature singularity, which must be included in the manifold because it is spacelike geodesically incomplete, this singularity is quite harmless for pointlike observers in the string frame. All black string manifolds turn out to be causally geodesically complete, since for arbitrary initial conditions at infinity all but one causal geodesics have a turning point before reaching the singularity. The one null geodesic which does not turn can come arbitrarily close to the singularity but can never reach it for a finite value of the affine parameter. Moreover, the first family possesses an interesting non-singular extremal limit, in that it has one hypersurface orthogonal null Killing vector, and nonvanishing gauge hair. We have employed this property to extend this solution to include a traveling wave, resorting to the generating technique proposed recently by D. Garfinkle in the context of string gravity[7]. The other family displays a peculiar combination of properties of both black strings and four dimensional black holes. Particularly interesting is the presence of the ergosphere, which arises entirely due to the axion and electric charges. This black string also has two different extremal limits. One of them corresponds to the wave-transformed extremal static solution . The other is reminiscent of the extremal Kerr-Newman case, and represents a gauge charged black string, with ergosphere but without null Killing vector.

We begin by outlining the essentials needed for the derivation of our new solutions. For simplicity we will restrict the discussion to three dimensions. Then, the effective action of string theory describing dynamics of massless bosonic background fields to the lowest order in the inverse string tension $\alpha'$ is, in the world-sheet frame,

$$S = \int d^3x \sqrt{g} e^{-\Phi} \left( R + \partial_\mu \Phi \partial^\mu \Phi - \frac{1}{12} H_{\mu\nu\lambda} H^{\mu\nu\lambda} - \frac{\alpha'}{4} F_{\mu\nu} F^{\mu\nu} + 2\Lambda \right). \qquad (1)$$

This action is written in Planck units $\kappa^2 = 1$. Here $F_{\mu\nu} = \partial_\mu A_\nu - \partial_\nu A_\mu$ is the field strength of the Abelian gauge field $A_\mu$, $H_{\mu\nu\lambda} = \partial_\lambda B_{\mu\nu} + \text{ cyclic permutations } - (\alpha'/2)\Omega_{M\mu\nu\lambda}$ is the field strength associated with the Kalb-Ramond field $B_{\mu\nu}$ and $\Phi$ is the dilaton field. The Maxwell Chern-Simons form $\Omega_{M\mu\nu\lambda} = A_\mu F_{\nu\lambda} + \text{ cyc. perm.}$ appears in the definition of the axion field strength due to the Green-Schwarz anomaly cancellation mechanism, and can be understood as a model-independent residue after dimensional reduction from ten-dimensional superstring theory. For convenience we will set $\alpha' = 1$.

We will be considering field configurations of the form

$$ds^2 = \Gamma(r) \, dr^2 + G_{jk}(r) \, dx^j dx^k,$$
$$B = \frac{1}{2} B_{jk}(r) dx^j \wedge dx^k, \quad A = A_j(r) dx^j, \quad \Phi = F(r), \qquad (2)$$

where $G_{jk}(r)$ is the metric of a 2 dimensional submanifold of signature 0. These backgrounds have two commuting Killing vector fields $\partial_k = \partial/\partial x^k$ and we can integrate them out. In this case the action (1) can be rewritten in the manifestly $O(2,3)$ invariant form

$$S_{eff} = \int dr \sqrt{\Gamma} e^{-\phi} \left( \frac{1}{\Gamma} \phi'^2 + \frac{1}{8\Gamma} \text{Tr}(\mathcal{M}' \mathcal{L})^2 + 2\Lambda \right), \qquad (3)$$

where the prime denotes the derivative with respect to $r$. Note that the physical dilaton $\Phi$ has been replaced by the effective dilaton $\phi = \Phi - (1/2)\ln|\det G|$ after dimensional reduction. Matrices $\mathcal{M}$ and $\mathcal{L}$ which appear in the action (3) are defined by

$$\mathcal{M} = \begin{pmatrix} g^{-1} & -g^{-1}C & -g^{-1}A \\ -C^T g^{-1} & g + a + C^T g^{-1} C & A + C^T g^{-1} A \\ -g^{-1} A & A + C^T g^{-1} A & 1 + A^T g^{-1} A \end{pmatrix}, \quad \mathcal{L} = \begin{pmatrix} 0 & \mathbf{1} & 0 \\ \mathbf{1} & 0 & 0 \\ 0 & 0 & 1 \end{pmatrix}. \quad (4)$$

Here $g$ and $b$ are $2 \times 2$ matrices defined by the dynamical degrees of freedom of the metric and the axion: $g = (G_{jk})$ and $b = (B_{jk})$. The column vector $A$ is built of the components of our gauge field. The matrices $a$ and $C$ are defined by $a = AA^T$ and $C = (1/2)a + b$ respectively, and $\mathbf{1}$ are two-dimensional unit matrices. Note that $\mathcal{M}^T = \mathcal{M}$ and $\mathcal{M}^{-1} = \mathcal{L}\mathcal{M}\mathcal{L}$ - thus we see that it is a symmetric element of $O(d, d+n)$. Consequently, a cogradient $O(d, d+n)$ rotation $\mathcal{M} \to \Omega \mathcal{M} \Omega^T$ is a symmetry of the action, and the equations of motion, because it represents a group motion which changes $\mathcal{M}$ while maintaining its symmetry property. The new $\mathcal{M}$ will give us the new solution.

We have applied this technique to two well-known solutions describing black strings. Starting from the black string family discovered by Horne and Horowitz[5], and using

$$\Omega = \begin{pmatrix} 1 & 0 & 0 & 0 & 0 \\ 0 & (1+c)/2 & 0 & (1-c)/2 & -s/\sqrt{2} \\ 0 & 0 & 1 & 0 & 0 \\ 0 & (1-c)/2 & 0 & (1+c)/2 & s/\sqrt{2} \\ 0 & -s/\sqrt{2} & 0 & s/\sqrt{2} & c \end{pmatrix}, \quad (5)$$

with $c = \cosh(2\alpha)$ and $s = -\sinh(2\alpha)$, we have obtained one of our new solutions:

$$ds^2 = \frac{dr^2}{2\Lambda(r - m\hat{c}^2)(r - m\hat{s}^2 - Q^2/m)} + \frac{r - m\hat{s}^2 - Q^2/m}{r - m\hat{s}^2} dx^2$$
$$- \frac{r - m\hat{c}^2}{r^2(r - m\hat{s}^2)}((r - m\hat{s}^2)dt - \hat{s}^2 Q dx)^2, \quad (6)$$

$$B = \frac{Q\hat{c}^2}{r} dx \wedge dt, \qquad A = -\sqrt{2}\frac{\hat{s}\hat{c}}{r}(Qdx + mdt), \qquad e^{-\Phi} = \sqrt{2\Lambda}r,$$

where $\hat{c} = \cosh\alpha$, $\hat{s} = -\sinh\alpha$. This solution generalizes the static family of Horne and Horowitz, reducing to it when $\alpha = 0$. It also has two horizons, the event horizon at $r = m\hat{c}^2$, and the Cauchy horizon at $r = m\hat{s}^2 + Q^2/m$.

We have found another solution by considering the electrically charged stringy 2D black hole[6] crossed with a flat line **R**. We have applied a different $O(2,3)$ twist to it, with $b$ a real number:

$$\Omega = \begin{pmatrix} 0 & 1 & b & 0 & 0 \\ 1 & 0 & 0 & -b & 0 \\ 0 & 0 & 0 & 1 & 0 \\ 0 & 0 & 1 & 0 & 0 \\ 0 & 0 & 0 & 0 & 1 \end{pmatrix}. \tag{7}$$

After some manipulations, involving coordinate transformations and introduction of ADM parameters representing the mass and the gauge charges, we have arrived at

$$ds^2 = \frac{dr^2}{2\Lambda(r-r_+)(r-r_-)} + (1 - \frac{\mathcal{Q}^2}{\mathcal{M}r} - \frac{e^2 \mathcal{Q}^2}{\mathcal{M}^2 r^2})dx^2$$
$$- (1 - \frac{\mathcal{M}}{r} + \frac{e^2}{r^2})dt^2 - 2\frac{\mathcal{Q}e^2}{\mathcal{M}r^2}dxdt, \tag{8}$$
$$B = \frac{\mathcal{Q}}{r}dx \wedge dt, \qquad A = -\sqrt{2}\frac{e}{r}(dt + \frac{\mathcal{Q}}{\mathcal{M}}dx), \qquad e^{-\Phi} = \sqrt{2\Lambda}r,$$

with the horizons now given by $r_\pm = (\mathcal{M}^2 + \mathcal{Q}^2 \pm \sqrt{\mathcal{M}^2 - \mathcal{Q}^2}\sqrt{\mathcal{M}^2 - \mathcal{Q}^2 - 4e^2})/2\mathcal{M}$. We should point out that similar variables for our first family of solutions give a representation far less transparent than the one provided by (8).

We will now present the causal structure of these solutions. Whereas some aspects of the geometry of these two solutions are remarkably similar, there are interesting differences. Both share some of the features of the static black string. They are both asymptotically flat configurations with two Killing fields, $\partial_x$ and $\partial_t$, with infinity described by the limit $\tilde{r} = \ln r \to \infty$, where they approach exponentially fast the linear dilaton vacuum, with flat Minkowski metric and vanishing gauge fields. As pointed above, both have an event and a Cauchy horizon. In addition, they have a scalar curvature singularity at $r = 0$.

More details can be found from investigating the geodesics of the solutions. Due to the presence of two Killing vector fields, the geodesic equations take a particularly simple form. For the general ansätz of equation (3) by introducing two integrals of motion associated with the cyclic coordinates $P_\mu = (-E, P)$, and the squared rest mass of the particle moving on the geodesic $p = 0, 1$ (distinguishing null and timelike geodesics), we obtain the following formula for the radial coordinate (the overdot denotes the derivative with respect to the affine parameter):

$$\Gamma r'^2 + G^{jk} P_j P_k = -p. \tag{9}$$

For brevity, we will only present the explicit form of this equation for the first of our solutions, given by (6). In this case, we can rearrange (9) to take the following form:

$$\frac{r'^2}{2\Lambda r^2} = E^2 - P^2 - p - \frac{(QE + mP)(QE - m(\hat{c}^2 + \hat{s}^2)P) - pm(\hat{c}^2 + \hat{s}^2)}{mr} \\ - \frac{(QE + mP)^2 \hat{s}^2 \hat{c}^2 + pm\hat{c}^2(m\hat{s}^2 + Q^2/m)}{r^2}. \tag{10}$$

This equation then governs the geodesic motion. We can now readily verify the above claims about the horizons etc. Here we will only review what happens in the region near the singularity. We can see that the coefficient of the $O(1/r^2)$ term in this equation is non-positive for all causal geodesics. As a consequence, no causal geodesic with this term being nonzero, beginning outside of the black string, can reach the singularity at $r = 0$, because the $O(1/r^2)$ term forces it to stop and turn. Thus, the only geodesics which don't turn away from the singularity at some finite value of $r$ are null geodesics with $QE + mP = 0$. However, in this case the equation (10) reduces to the standard linear homogenous equation, with exponential solutions $r \propto \exp(\lambda)$, and so for any finite $\lambda$, $r \neq 0$. As a consequence, our spacetime is timelike *and* null geodesically complete. The singularity still must be included in the manifold, which is spacelike geodesically incomplete. In addition, it can also be reached by nongeodesic causal curves, or possibly by causal observers in different conformal frames. Yet, it is remarkable that in the string frame, pointlike observers living serenely along causal geodesics can never reach it. Repeating this analysis for the solution (8) we can see that the singularity there also has this property.

At this point, we can give the Penrose diagram describing the causal structure of our solutions, but with the following warning. We should note that the behaviour of the coordinates while crossing these surfaces is somewhat different from the situation enjoyed by the Horne-Horowitz black string. While the radial coordinate behaves the same, being spacelike outside the event horizon and inside the Cauchy horizon, and timelike in between, the time at infinity $t$, which turns spacelike after crossing the event horizon, regains the timelike character again after crossing the interior static limit inside the Cauchy horizons. Likewise, $x$ also changes signature, becoming timelike crossing its Killing horizons. As we see, there are regions where the time coordinate is an $r$-dependent linear combination of $t$ and $x$. Keeping this in mind, we can still employ the diagrammatic technique as a descriptive tool. The Penrose diagram thus is

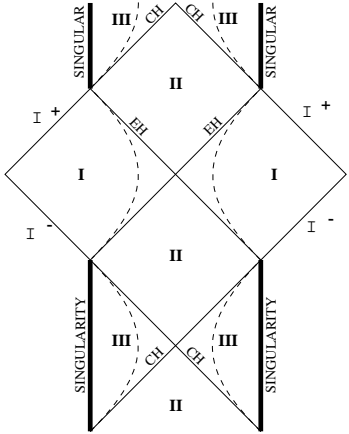

Figure 1.

The dashed line in the region I is the ergoregion and exists only in the solution (8). The other dashed line, in region III, is the interior static limit and is present in both (6) and (8).

Extremal limits in our case are defined as the conditions for the two horizons to coincide. This gives $|Q| = m$ for solution (6) and $\mathcal{Q}^2 = \mathcal{M}^2$ or $\mathcal{Q}^2 + 4e^2 = \mathcal{M}^2$ for solution (8). This is in agreement with the investigation of the Hawking temperatures for the two solutions. Employing the approach designed for stationary configurations[8], we readily obtain for the temperatures

$$T = \sqrt{\frac{\Lambda}{2}} \frac{\sqrt{m^2 - Q^2}}{2\pi m \hat{c}^2}, \tag{11}$$

for the solution (6), and

$$T = \sqrt{\frac{\Lambda}{2}} \frac{\sqrt{\mathcal{M}^2 - \mathcal{Q}^2}\sqrt{\mathcal{M}^2 - \mathcal{Q}^2 - 4e^2}}{\sqrt{\mathcal{M}^2 - \mathcal{Q}^2} + \sqrt{\mathcal{M}^2 - \mathcal{Q}^2 - 4e^2}}, \quad (12)$$

for (8). In both cases, the temperature vanishes in the extremal limit. Thus the interpretation of these configurations as extremal black strings.

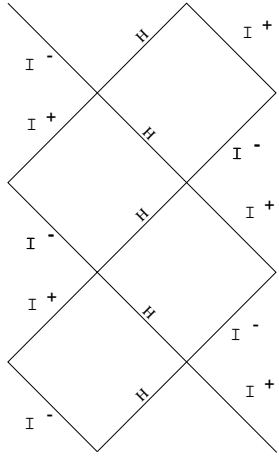
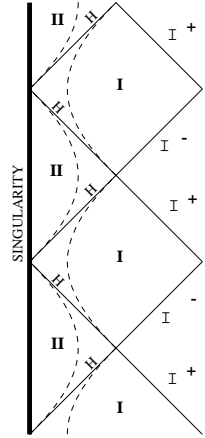

Figure 2a.          Figure 2b.

There are considerable differences between these extremal limits. Naively, we would expect to obtain a solution with a singularity enclosed by a single horizon. Whereas this is true for the extremal limit of (8), when we look at the extremal limit of the first family (6) we can see that, like in the static case, the coordinate $r$ does not give the proper extension across the horizon $r = m\hat{c}^2$. In this coordinate, it appears that no timelike or null geodesics can cross the horizon. The proper extension is given by the new radial coordinate $\bar{r}^2 = r - m\hat{c}^2$. Horizon is now located at $\hat{r} = 0$, and the metric can be extended beyond it, to the region $\hat{r} < 0$. As the solution is invariant under the radial reflection $\hat{r} \to -\hat{r}$, we see that a particle finds itself in a universe identical to the one which it just left. As a consequence, the maximal extension of this solution is based on a zig-zag event horizon along which an infinite number of asymptotically flat regions are connected, with the causal structure identical to the extremal black string of Horne and Horowitz (Fig. 2a). In contrast, the causal structure of the extremal limit $\mathcal{Q}^2 + 4e^2 = \mathcal{M}^2$ is more similar to the extremal Reissner-Nordström black hole, except that here we see the ergosphere (denoted in Fig. 2b. by the dashed line in region I). The remaining extremal limit $\mathcal{Q}^2 = \mathcal{M}^2$ in fact is a completely nonsingular geometry, because all geodesics turn before reaching the event horizon, and so the "interior" is excluded from the manifold.

Another interesting property of the first extremal limit is that it possesses a null Killing vector. This allows us to generalize the solution to include a traveling wave propagating along the string, using the wave generating technique[7]. We will only give the solution here, which takes the form:

$$ds^2 = \frac{dr^2}{2\Lambda(r - m\hat{c}^2)^2} - 2(1 - \frac{m\hat{c}^2}{r})d\chi d\tau + \left((1 - \frac{m\hat{c}^2}{r})\frac{m\hat{s}^2}{r} + \frac{2\Lambda}{r}f(\tau)\right)d\tau^2. \qquad (13)$$

The matter sector remains the same as in the extremal limit of (6). The arbitrary function $f(\tau)$ determines the shape of the wave profile. We note that it also has a considerable effect on the global structure of the space-time. If we restrict out attention to the case when $f = $ const., we note that if $f = -ms^2/2\Lambda$ the solution is the same as the extremal limit $\mathcal{Q}^2 = \mathcal{M}^2$ of (8). More generally, we see the following three possibilities: (i) $f > 0$ solution globally resembles the extremal limit $\mathcal{Q}^2 + 4e^2 = \mathcal{M}^2$, but as it has a null Killing vector, there is no ergosphere; (ii) $f = 0$ is our starting solution, globally defined by Fig. 2a; (iii) $f < 0$ is of the same structure as the $\mathcal{Q}^2 = \mathcal{M}^2$ limit of (8). For all these solutions the Hawking temperature vanishes identically and thus they can be interpreted as various extremal black strings.

In summary, we have presented two new, gauge-charged, black strings as well as a plethora of extremal limits of these solutions. As we do not expect the violations of the spirit of no-hair theorems in the theory defined by the action (1), we conjecture that there must exist a more general family of solutions. The extra parameter representing the linear momentum along the string would guarantee the existence of continuous limits towards the extremal solutions we have encountered. In addition, as the family (8) has an ergoregion outside of the event horizon, there exists the possibility for energy extraction by Penrose processes. The resulting effect would normally be the shrinking of the ergoregion, which in this case is supported solely by the gauge charges. Thus this would violate charge conservation. The way out of this problem is again the existence of the more general family of solutions. The shrinking of the ergoregion would be at the expense of loosing the momentum along the string, i.e. the energy loss should "boost" the string by $-P$. The general solution thus would be an interesting example to look at the energy balance.

## Acknowledgments

The author would like to thank W.G. Anderson for a fruitful collaboration. Thanks are also due to B. Campbell, W. Israel, R. Myers, R. Madden and E. Martinez for useful discussions. This work has been supported in part by the National Science and Engineering Research Council of Canada, and in part by an NSERC postdoctoral fellowship.